\documentclass[journal]{IEEEtran}

\usepackage{xcolor,soul,framed} 
\usepackage{physics}

\colorlet{shadecolor}{yellow}
\usepackage[pdftex]{graphicx}
\graphicspath{{../pdf/}{../jpeg/}}
\DeclareGraphicsExtensions{.pdf,.jpeg,.png}

\begin{document}

\bstctlcite{IEEEexample:BSTcontrol}
    \title{Entangled Photon Pair Source Demonstrator using the Quantum Instrumentation Control Kit System}
     

\author{Si Xie, Leandro Stefanazzi, Christina Wang, Cristi\'{a}n Pe\~{n}a, Raju Valivarthi, Lautaro Narv\'{a}ez, Gustavo Cancelo, Keshav Kapoor, Boris Korzh, Matthew Shaw, Panagiotis Spentzouris, and Maria Spiropulu
\thanks{G. Cancelo, K. Kapoor, C. Pe\~{n}a, L. Stefanazzi, and P. Spentzouris are with the Fermi National Accelerator Laboratory, Batavia, IL 60510, USA}%
\thanks{S. Xie is with the Fermi National Accelerator Laboratory, Batavia, IL 60510, USA, the Division of Physics, Mathematics and Astronomy and Alliance for Quantum Technologies (AQT), California Institute of Technology, Pasadena, CA 91125, USA}%
\thanks{C. Wang, R. Valivarthi, L. Narv\`{a}ez, and M. Spiropulu are with the Division of Physics, Mathematics and Astronomy and Alliance for Quantum Technologies (AQT), California Institute of Technology, Pasadena, CA 91125, USA}%
\thanks{B. Korzh, and M. Shaw are with the Jet Propulsion Laboratory, California Institute of Technology, Pasadena, CA 91109, USA}%

}



\maketitle

\begin{abstract}
We report the first demonstration of using the Quantum Instrumentation and Control Kit (QICK) system on RFSoC-FPGA technology to drive an entangled photon pair source and to detect the photon signals.
With the QICK system, we achieve high levels of performance metrics including coincidence-to-accidental ratio exceeding 150, and entanglement visibility exceeding 95\%, consistent with performance metrics achieved using conventional waveform generators. 
We also demonstrate simultaneous detector readout using the digitization functional of QICK, achieving internal system synchronization time resolution of 3.2 ps.
The work reported in this paper represents an explicit demonstration of the feasibility for replacing commercial waveform generators and time taggers with RFSoC-FPGA technology in the operation of a quantum network, representing a cost reduction of more than an order of magnitude. 
\end{abstract}

\begin{IEEEkeywords}
quantum network, quantum communication, fiber optics, photon pair, C-band, FPGA, control electronics,
\end{IEEEkeywords}

%
\IEEEpeerreviewmaketitle

\section{Introduction}


Quantum networks hold great promise for revolutionizing the way we communicate and process information~\cite{qnetwork,IEQNET}. 
They can offer unparalleled security; enable new fundamental scientific discoveries through networks of quantum sensors and quantum computers; serve critical practical solutions applications in industry such as in finance, supply chain management, or cybersecurity; and drive technological innovation and economic growth through the creation of new industries and markets. 

Field Programmable Gate Arrays (FPGAs) are highly configurable hardware devices that can be programmed to perform specific tasks, making them well-suited for implementing quantum communication protocols. 
By using FPGAs, researchers and developers can easily reconfigure their systems to support different types of quantum operations, allowing them to rapidly iterate and improve upon their designs.
FPGAs can also help improve the efficiency of quantum communication systems by offloading certain tasks from the main processor, reducing the overall cost and complexity of the system. 
In addition, FPGAs can support the scalability of quantum communication systems by allowing them to be easily deployed,  reconfigured, and upgraded as needed.

System specific control and readout increase the functionality of quantum systems, integrate complex functions, and eliminate the bottlenecks and synchronization problems of using expensive off-the-shelf controls. 
In this paper we present results of a photon entanglement distribution experiment using accurate pulse generation and readout based on an FPGA with integrated analog-to-digital and digital-to-analog conversion. 
The control electronics is based on the open Quantum Instrumentation and Control Kit (QICK) system~\cite{QICK}, developed at Fermilab and widely adopted for control of superconducting qubit experiments. 
The QICK was expanded to provide the functionality required for Quantum Networks. 
The board used in this experiment was the ZCU216, as it provides 10~Gsps digital-to-analog converters (DACs) and $2.5$~Gsps analog-to-digital converters (ADCs).

Experimental demonstration are shown using this control system for a quantum network by producing entangled photon-pairs and measuring its entanglement quality.
Through this demonstration, we show the feasibility in realizing quantum networks with nodes that are fully controllable with a single FPGA, allowing for an economical and easily deployable solution to scalable quantum networks.

\section{QICK for Quantum Networks}
\label{sec:fpga}
\begin{figure}
  \begin{center}
  \includegraphics[width=3.5in]{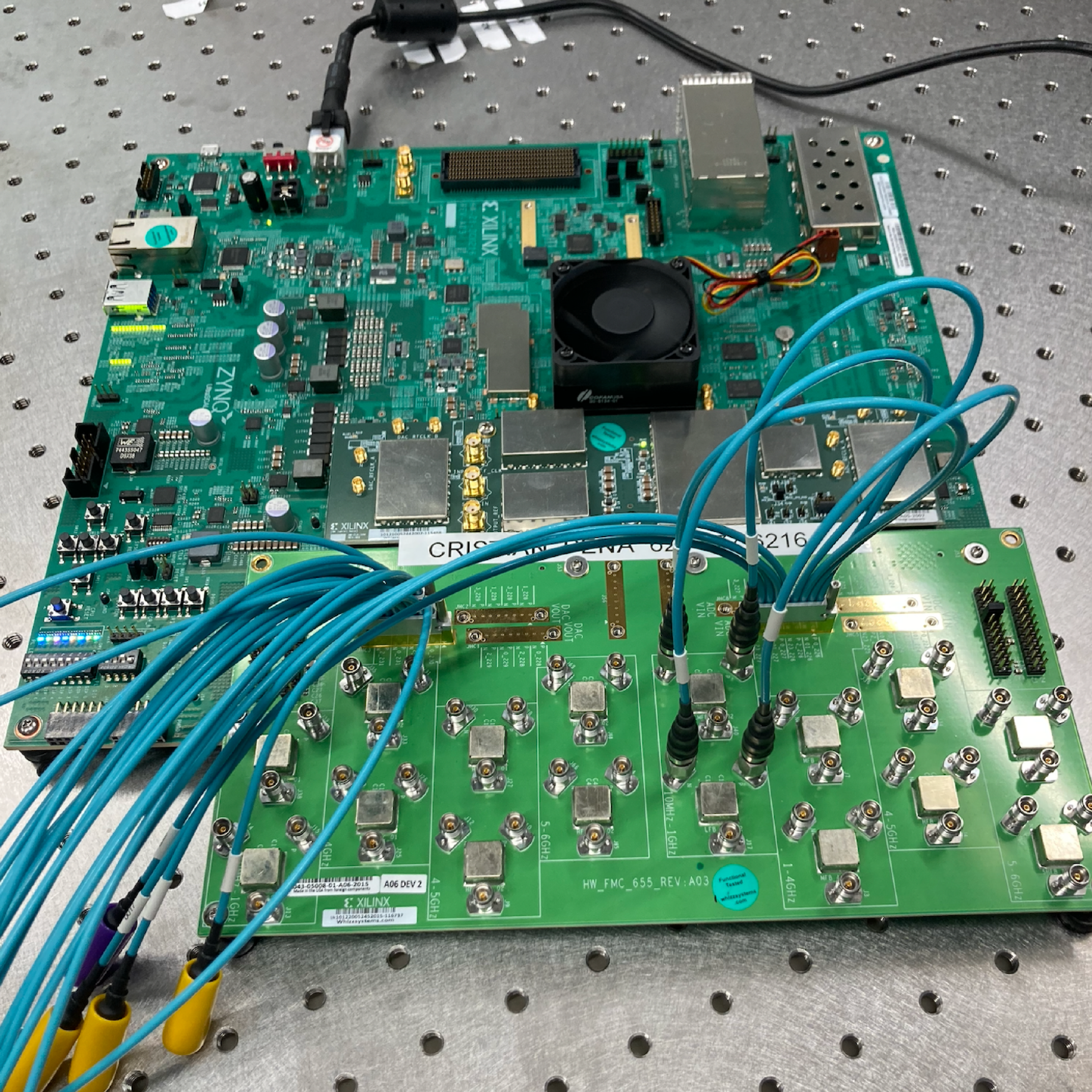}
  \caption{A photograph of the QICK System based on Xilinx's ZCU216.}
  \label{fig:FPGA_photo}
  \end{center}
\end{figure}

QICK was originally developed for Superconducting Qubit experiments~\cite{QICK}. 
The first version was deployed over the Xilinx ZCU111 development board. 
Later it was extended to fully support ZCU216 and RFSoC 4x2 boards. 
The ZCU216 version of the QICK was selected for this experiment. 
This board features a Xilinx RFSoC generation 3 UltraScale+ device, and it can provide up to 16 output DACs at a speed of 10~Gsps, and up to 16 input ADCs at $2.5$~Gsps, which makes this platform ideal for Quantum Networks applications. 
Figure~\ref{fig:FPGA_photo} shows a photograph of the ZCU216 connected to the experiment.

The requirements for the control systems of quantum networks are different from those of superconducting qubit experiments. For quantum networks, very fast and short pulses need to be generated to drive the electro-optic intensity modulators used to create time-bin entangled photon pairs that are ideal for long distance fiber-based quantum networks. 
For this application, pulses that are 100--200~ps wide are produced, with the "early" and "late" states in the time-bin basis, separated by a time delay of about 1--2~ns. The repetition rate is set to be 100~MHz.

\begin{figure}
  \begin{center}
  \includegraphics[width=3.5in]{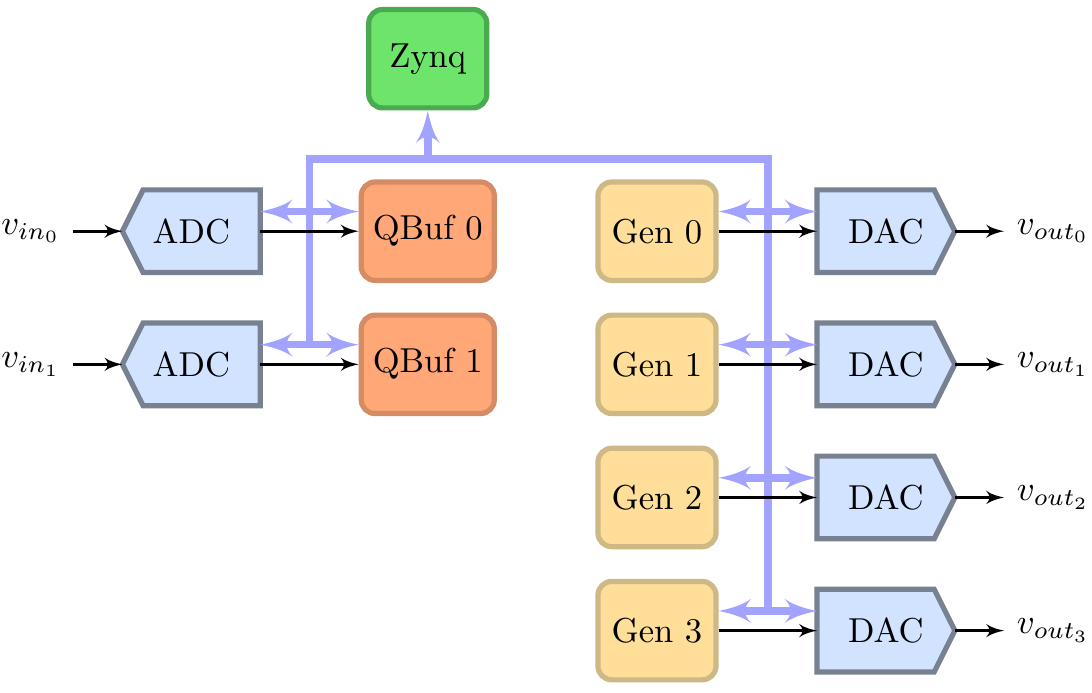}
  \caption{Block diagram of the QICK firmware used for this quantum network demonstrator experiment.}
  \label{fig:FPGA_firmware}
  \end{center}
\end{figure}%
The block diagram of the QICK system customized for this experiment can be seen on Fig.~\ref{fig:FPGA_firmware}.
To create the output pulses, a custom block was designed that allows to specify up to 32 arbitrary amplitude samples. 
These samples are read circularly and sent out by driving one DAC channel. 
The block can be easily replicated to populate all 16 available DAC channels. 
As it can be seen in Fig.~\ref{fig:FPGA_firmware}, four instances of the Pulse Generator Block were added connecting to four independent DACs. 
The sampling frequency of the output DACs for this experiment was set to $f_s=8.1$~Gsps, which gives about 123~ps per sample. 
Although the frequency could have been increased to 10~Gsps, the final value was chosen to match the pulse separation constraint of the external interferometer. 
The repetition rate of the 32-samples pattern is programmable from 0 to $2^{32}$ clock ticks, which corresponds to a maximum time of $8.4$~s. 
A value of 0 means the pattern is repeated without any gaps. 
Additional wait times can be added to slow the repetition rate.

On the detection side, superconducting nanowire single photon detectors (SNSPDs)~\cite{Marsili2013} produce signal pulses with amplitude between 500 and 900~mV, which are then sent to the ADC input channels of the QICK system. 
The sampling frequency of the ADC was set to its maximum of 2.5~Gsps. 
Figure~\ref{fig:FPGA_firmware} shows the two Qualifier Buffer (QBuf) blocks used for capture, connected to their corresponding ADC. 
To lower the FPGA memory requirements, pulses are captured by the QBuf block only when the input signal exceeds a threshold value. 
The threshold and capture window length are both configurable from the QICK software interface. 
The buffer block adds a time-tag to allow the measurement of signal pulses correlated across different channels. 
Once the signal pulses are captured, the user can retrieve the buffer information with all the pulses and time-tags for further offline processing. 
These offline processing routines can be easily implemented on the FPGA ARM processors to allow for continuous and real-time event count and tagging, an important next step towards operating scalable quantum networks that we leave for future work.

\section{Experimental Setup}

Using a standard entangled photon-pair source, we constructed a simple demonstrator experiment illustrating the use of the RFSOC-FPGA in photonic time-bin encoded quantum networks. 
The same source has been used for past work ranging from demonstration of high fidelity quantum teleportation~\cite{valivarthi2020teleportation} to demonstration of picosecond precision time synchronization~\cite{Valivarthi:2022vni,FNALANLSync}. 
Light at 1536~nm wavelength produced by a continuous wave fiber-coupled laser is directed into a fiber-coupled Mach-Zehnder Modulator (MZM)~\cite{ixblue} to produce pulsed light. 
The MZM is driven by radio-frequency (RF) pulses generated either by a commercial arbitrary waveform generator (AWG) or by the DAC functionality of the RFSOC-FPGA, and subsequently amplified to achieve the maximum extinction ratio of the MZM. 
The DAC sampling rate of our RFSOC-FPGA is 10~Gsps, resulting in the shortest possible pulses with a width of 100~ps. 
Due to bandwidth limitation on our RF amplifier, we create pulses with a width of about 200~ps, resulting in amplified RF pulses with a width of about 250~ps at half maximum, as shown in Figure~\ref{fig:FPGAPulses}.
The pulsed light is directed into an erbium-doped fiber amplifier (EDFA) and then sent through a periodically poled lithium niobate (PPLN) waveguide, upconverting the 1536~nm light to 768~nm.
A band-pass filter is used to remove residual 1536~nm light.
Time-correlated photon pairs at the original wavelength of 1536~nm are produced through the Type-II spontaneous parametric down conversion process (SPDC) in a second PPLN waveguide receiving the 768~nm light as input.

\begin{figure}
  \begin{center}
  \includegraphics[width=3.5in]{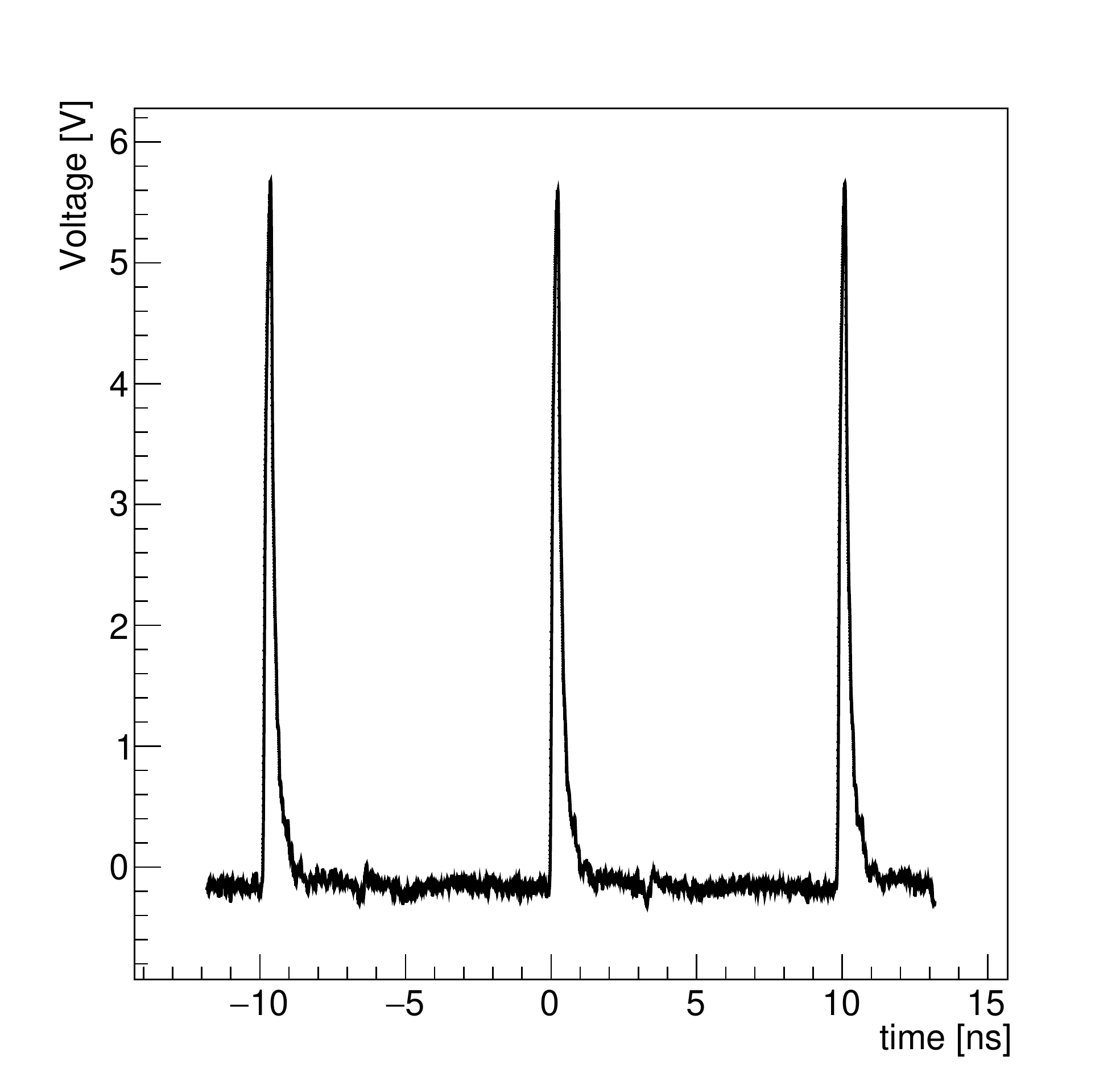}
  \includegraphics[width=3.5in]{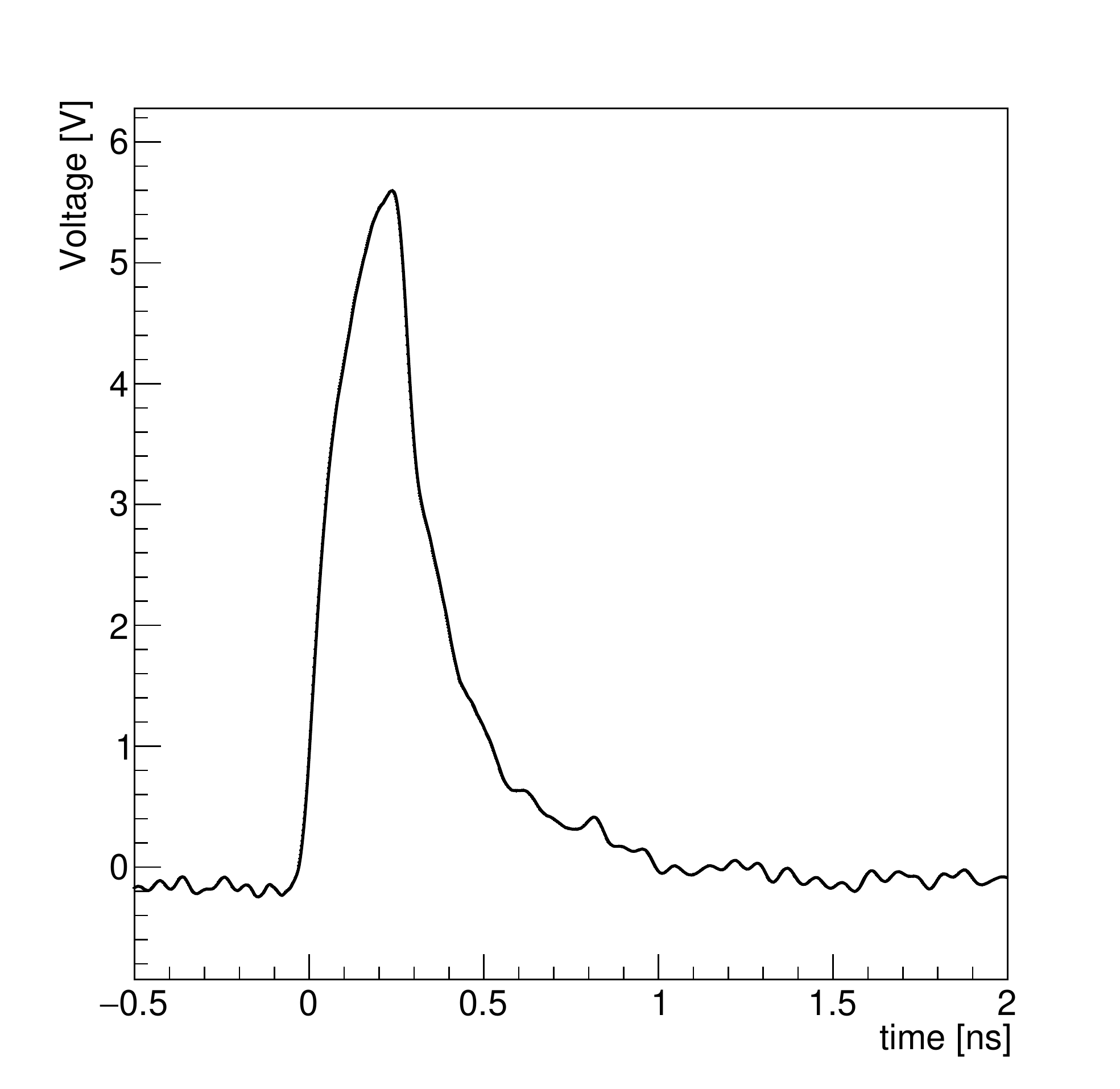}\\
  \caption{ Oscilloscope traces of the pulses generated by the RFSoC FPGA DAC after amplification. On the top, we show three repetitions of the same pulse structure with a rep rate of 100~MHz. On the bottom, we show a zoomed in version of a single repetition of the pulse.  }
  \label{fig:FPGAPulses}
  \end{center}
\end{figure}

\begin{figure}
  \begin{center}
  \includegraphics[width=3.5in]{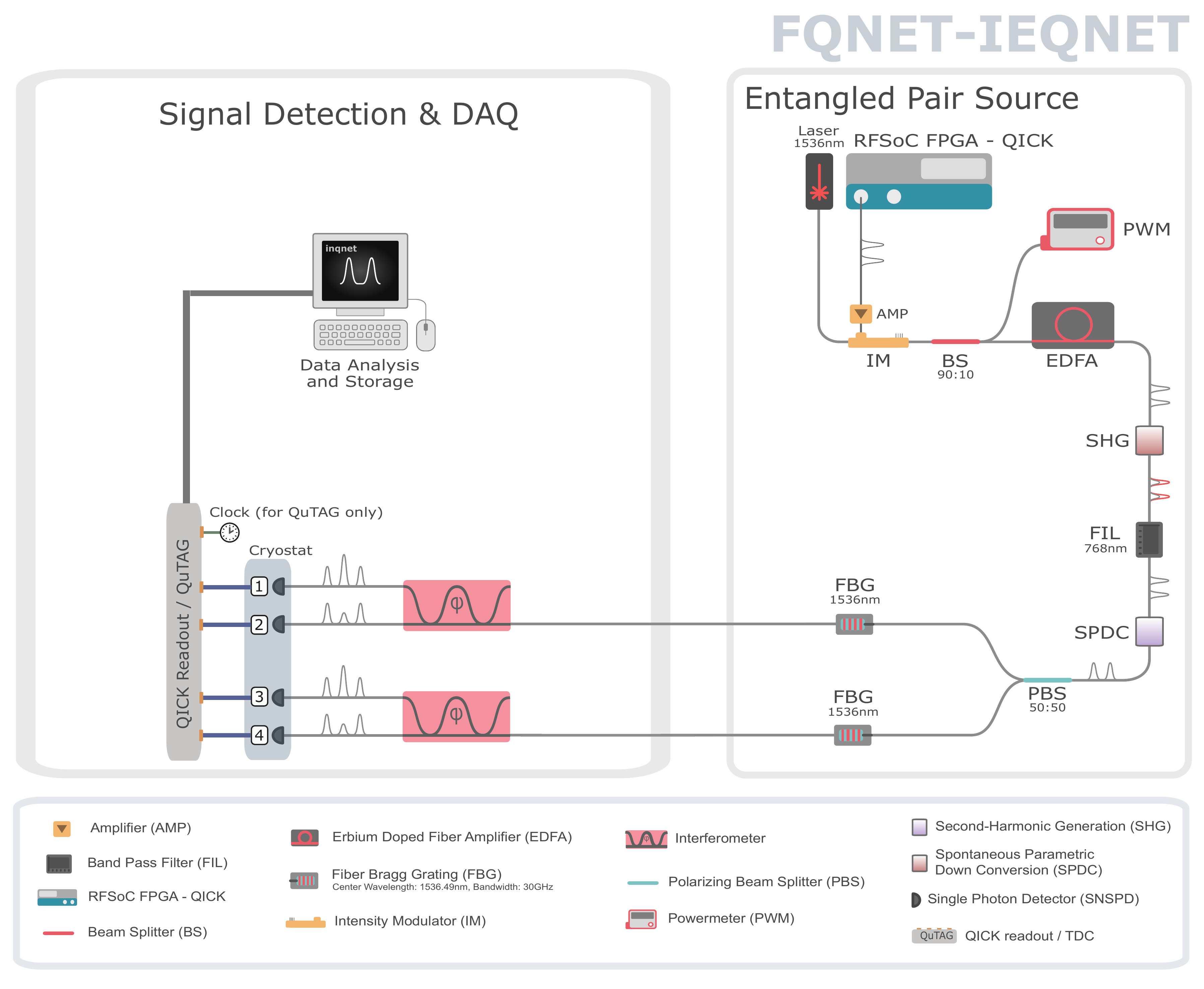}
  \caption{ Schematic diagram of the entangled photon-pair source setup used to characterize the RFSOC-FPGA QICK functionality. 
  \label{fig:setup}
    }
  \label{fig:FPGAPulses}
  \end{center}
\end{figure}

A fiber-based polarizing beam splitter separates the resulting pair of photons and both are directed into different SNSPD detectors.
The SNSPD signals are either time-tagged by a commercial time-to-digital (TDC) converter, or digitized by the ADC functionality of the RFSOC-FPGA. 
In both cases, we build coincidences in the detection of both photons, and record the differences in the detection time of the two photons. 
The coincidence-to-accidental (CAR) ratio is used to characterize and compare the quality of the entangled photon-pair source driven by the more conventional AWG and by the RFSOC-FPGA. 
To further characterize the quality of the entangled photon-pair source, we also use a Michelson interferometer to measure the entanglement visibility of the entangled photon-pair source in the x-basis driven by the RFSOC-FPGA.
Finally, we demonstrate the use of the RFSOC-FPGA in the detection of the photon signals by comparing the CAR measured using the commercial TDC and the one measured using the custom digitizer firmware and pulse-shape reconstruction software.
The experimental setup described above is shown schematically in Figure~\ref{fig:setup}

\section{Results}
We show two sets of results to characterize the performance of the pulse generation and signal readout functions of the RFSOC FPGA, respectively. 

\subsection{Pulse Generation}
\label{sec:ResultsPulseGeneration}
\begin{figure}
  \begin{center}
  \includegraphics[width=3.5in]{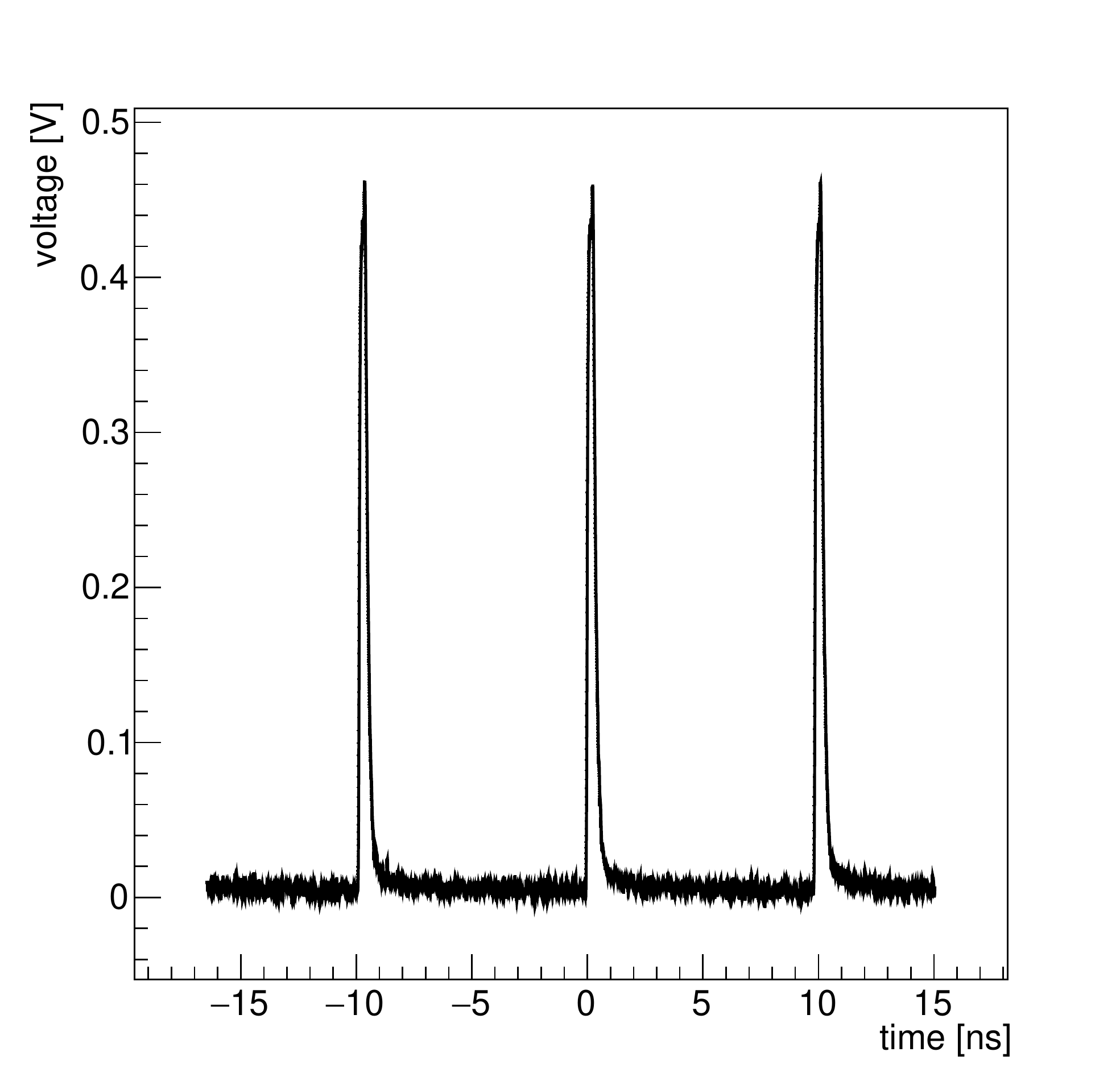}\\
  \caption{ Oscilloscope trace of the optical pulses at the output of the EDFA amplifier as measured by a 30~GHz InGaAs photodetector.  }
  \label{fig:ExtinctionRatio}
  \end{center}
\end{figure}

For the pulse generation functionality, we use the RFSOC-FPGA to drive the MZM.
The time structure of the pulsed light at the output of the EDFA is shown in Figure~\ref{fig:ExtinctionRatio} as measured by a 30~GHz InGaAs photodetector, showing an excellent MZM extinction ratio exceeding 20~db. 
The MZM driven by a commercial AWG achieved the same extinction ratio.

\begin{figure}
  \begin{center}
  \includegraphics[width=3.5in]{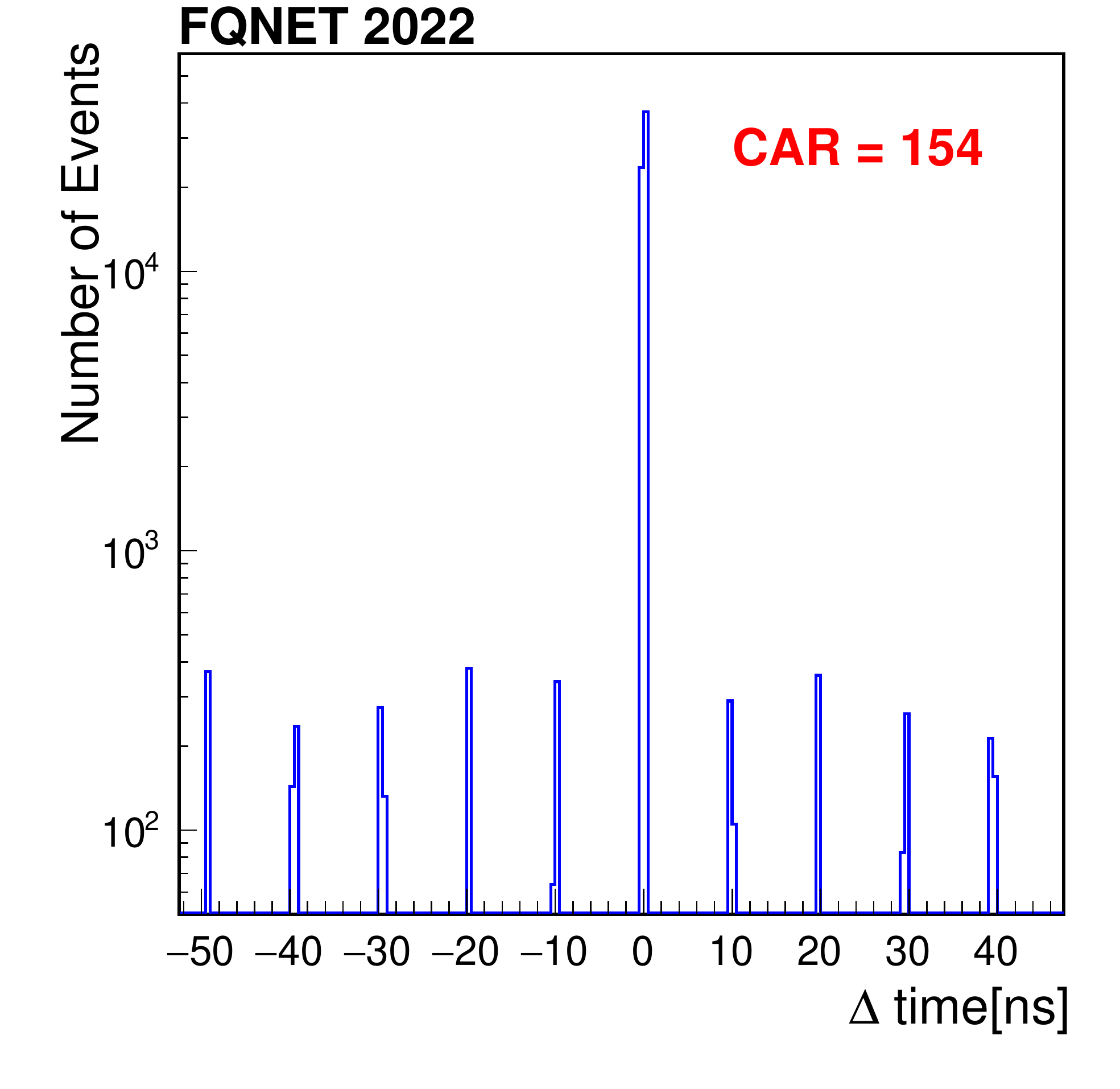}\\
  \caption{Histogram of the time difference between the photon pairs detected by two SNSPD detectors. From the histogram we measure a CAR of 154.}
  \label{fig:CARFromDaisy}
  \end{center}
\end{figure}

Using the entangled photon pairs produced by the pulsed light from the MZM driven by the RFSoC-FPGA, we measure the CAR to quantify the quality of the entangled photon-pair source.
The time difference between the photons in the two detectors is shown in Figure~\ref{fig:CARFromDaisy} as measured using a commercial TDC device, from which we measure a CAR of 154.

\begin{figure}
  \begin{center}
  \includegraphics[width=3.5in]{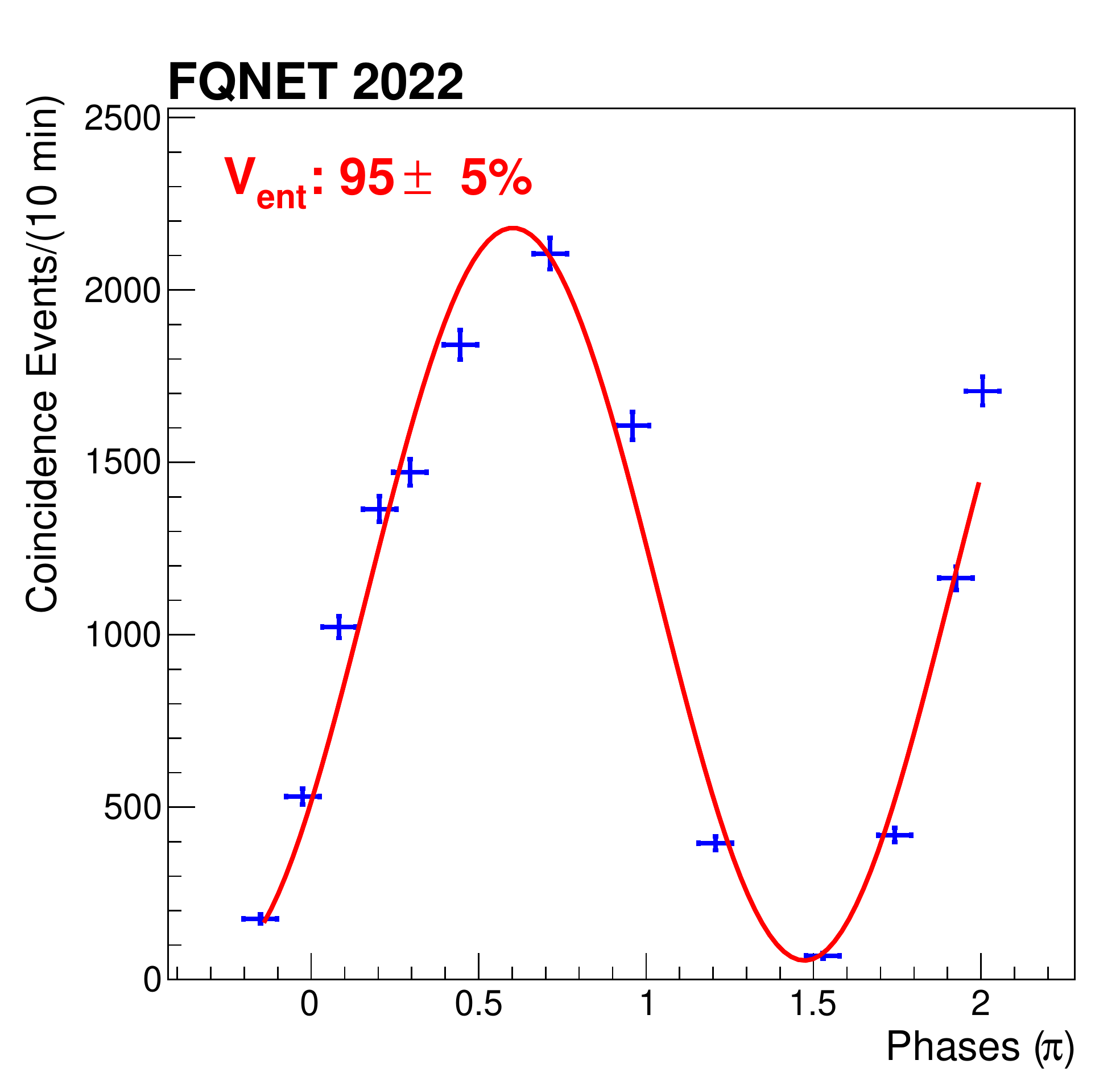}\\
  \caption{Coincidence counts on the outputs of the Michelson interferometers for each of the photon pairs of the entangled photon pair source is shown as a function of the scanned phase of one of the interferometers. This measurement uses the x-basis time-bin state ($\ket{\mathrm{early}} + \ket{\mathrm{late}}$).}
  \label{fig:EntanglementVisibility}
  \end{center}
\end{figure}

Furthermore, we produce time-bin entanglement  $\ket{\Psi} = \frac{\ket{\mathrm{ee}} + \ket{\mathrm{ll}}}{\sqrt{2}}$, and we measure the entanglement visibility using two fiber-based Michelson interferometers. 
In Figure~\ref{fig:EntanglementVisibility}, we show the coincidence counts as a function of the relative phase difference between the interferometers.
We perform a fit of the data to a sinusoidal function and measure the entanglement visibility to be $95\pm5\%$, indicating high quality entanglement. 

\subsection{Signal Readout}
\begin{figure}
  \begin{center}
  \includegraphics[width=3.5in]{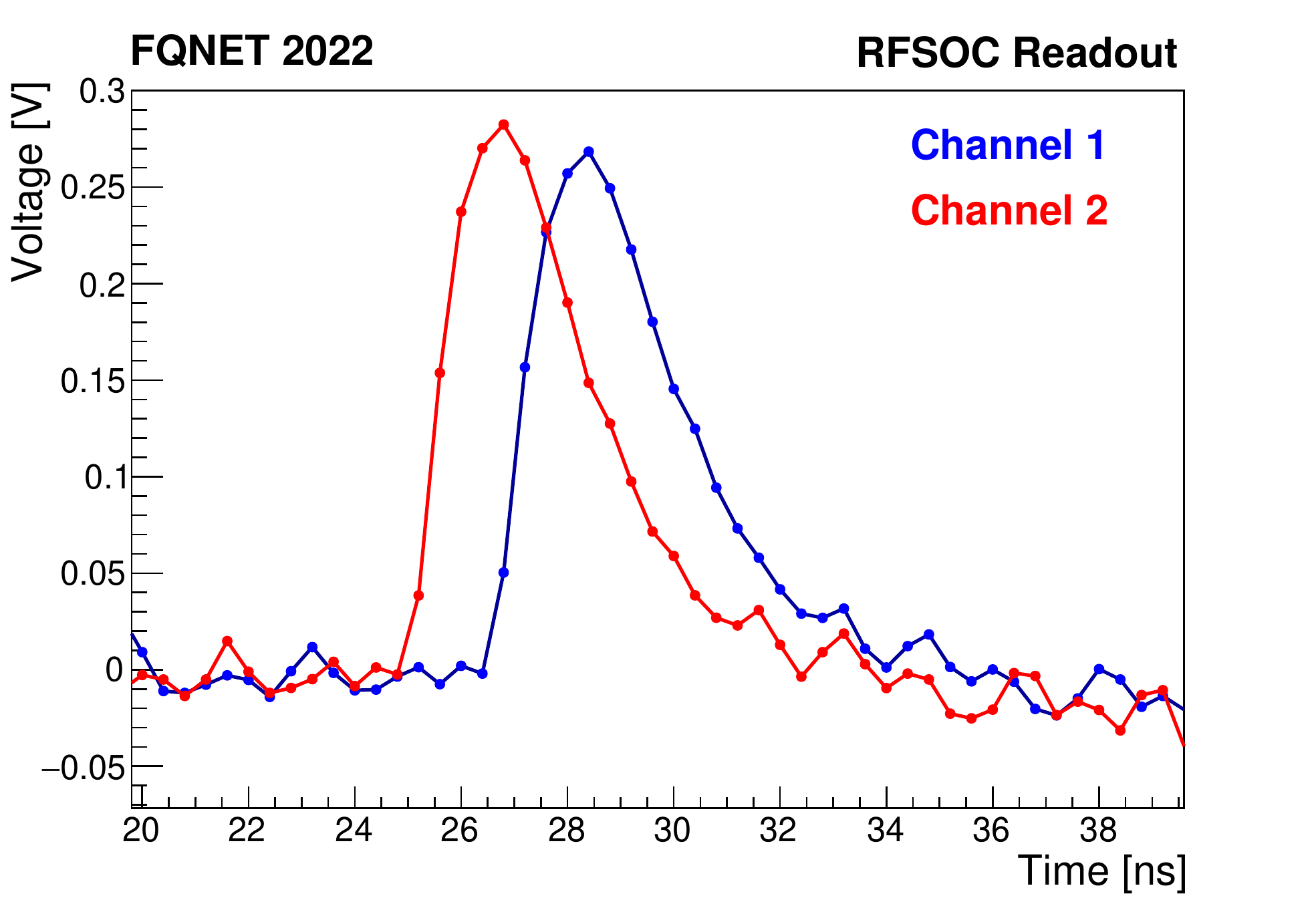}\\
  \caption{Pulse Read out by FPGA }
  \label{fig:FPGAReadoutPulse}
  \end{center}
\end{figure}

To demonstrate the signal readout capability of the RFSoC-FGPA, we repeat the CAR measurement described above in Section~\ref{sec:ResultsPulseGeneration} using the ADC of the RFSoC-FPGA to digitize the signal pulses.
An example of the SNSPD signal waveforms digitized by the RFSoC-FPGA ADCs are are shown in Figure~\ref{fig:FPGAReadoutPulse}, where the signals from the two detectors have been intentionally separated by 2~ns.

\begin{figure}
  \begin{center}
  \includegraphics[width=3.5in]{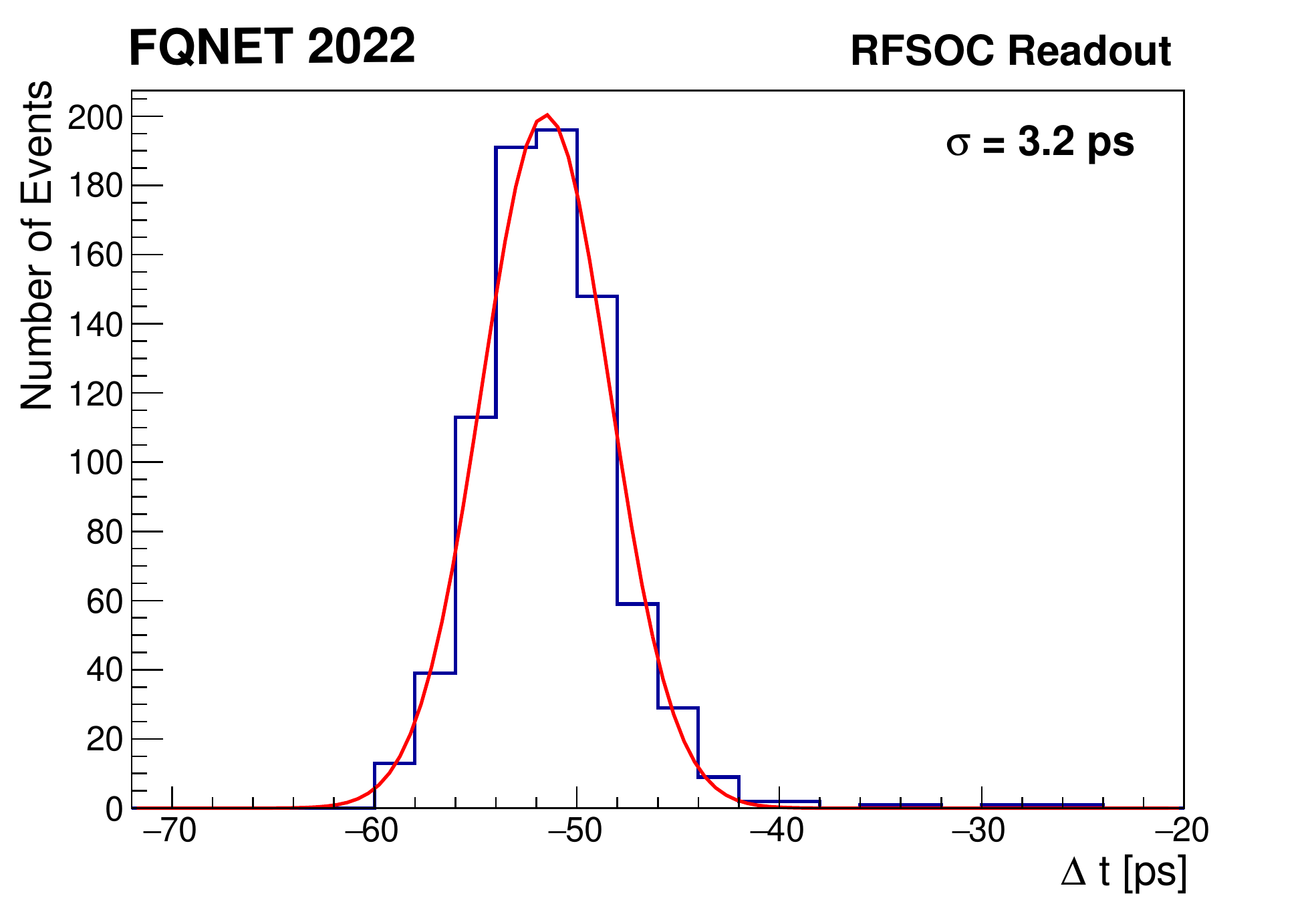}\\
  \caption{FPGA board resolution }
  \label{fig:FPGATimeResolution}
  \end{center}
\end{figure}

We first measure the time resolution introduced by the RFSOC-FPGA.
We split the RF signal from a single SNSPD detector and connect them to two different input channels on the RFSOC-FPGA.
We measure the time difference between the two channels over an ensemble of photon detection events and obtain the histogram shown in Figure~\ref{fig:FPGATimeResolution}.
We fit the histogram to a Gaussian function and obtain a time resolution of 3.2~ps. 

\begin{figure}
  \begin{center}
  \includegraphics[width=3.5in]{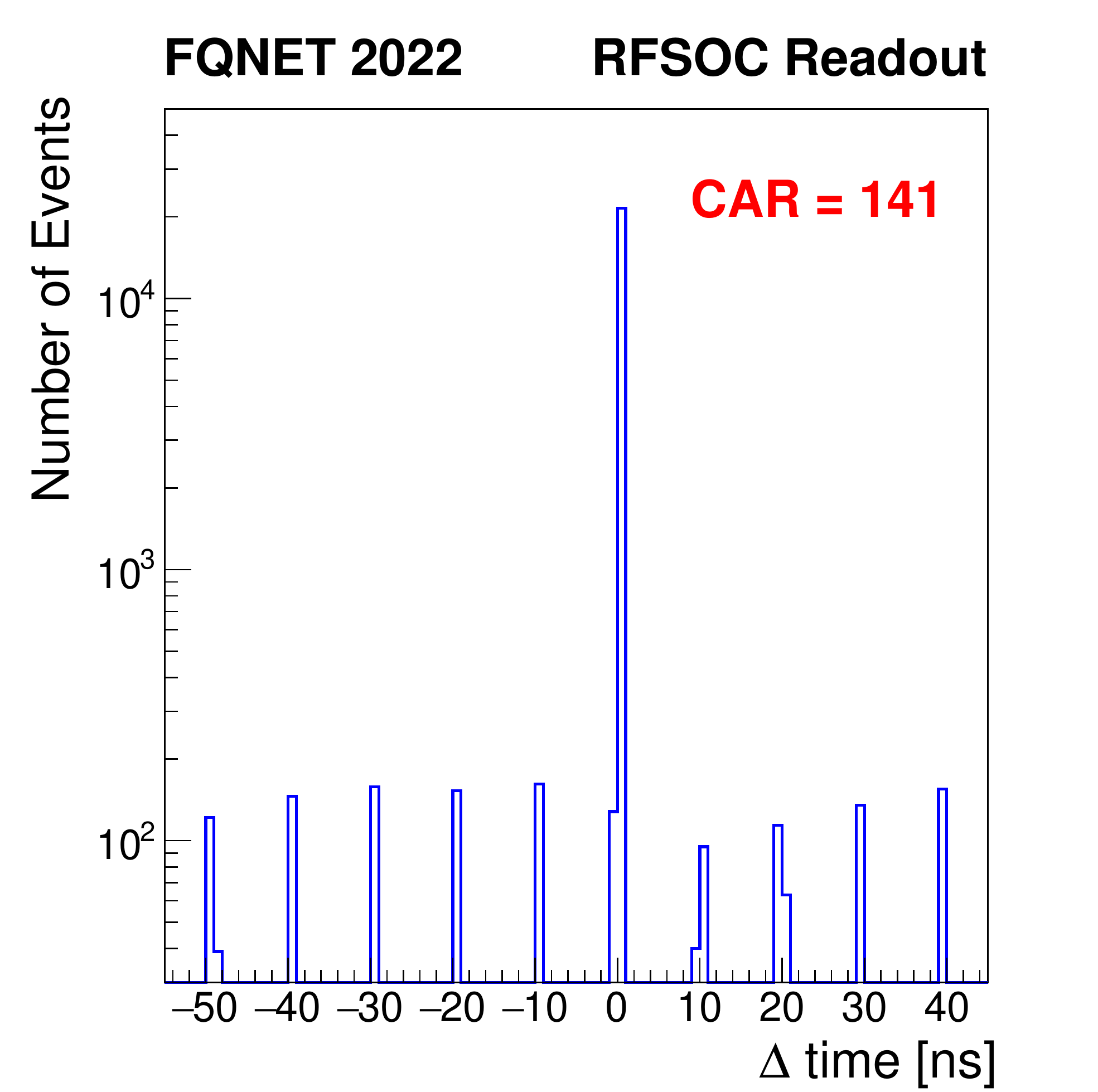}\\
  \caption{CAR plot with FPGA Readout. }
  \label{fig:CAR_FPGAReadout}
  \end{center}
\end{figure}

Finally, we use the waveform digitizer readout functionality of the RFSOC-FPGA to record pulses from two independent SNSPD detectors.
We build coincidence events by searching for any pair of pulses detected within a window of 100~ns, and obtain the time difference histogram in Figure~\ref{fig:CAR_FPGAReadout}.
From this histogram we measure the CAR to be 141, which compares well with a CAR of 154 measured using the commercial TDC readout system. 

\section{Discussion and outlook}
\label{sec:discussion}

The results reported in this paper represent the first feasibility demonstration for using the RFSoC-FPGA technology in key components of an operational quantum network.
Using the RFSoC-FPGA equipped with our custom firmware in lieu of the more conventional commercial AWG and time taggers, we achieved the same levels of performance metrics including the coincidence-to-accidental (CAR) ratio, the entanglement visibility, and the cross-channel signal detection time resolution. 
This work represents a practical demonstration of the power and importance of the RFSoC-FPGA technology in future quantum network design and implementation.
As commercial AWG's and time taggers cost on the order of 150K and 50K, respectively, in 2022 US dollars, while current state of the art RFSoC-FPGA's cost about 13K US dollars each, there is a potential for a 15 times reduction in the cost of the high speed electronics equipment necessary for operating quantum networks. 
Thus, the RFSoC-FPGA technology will no doubt play a key role in the deployment of scalable quantum networks. 

The next steps for RFSOC-FPGA development includes developing and optimizing the online time-tagging functionality, customizing the noise filters on the signal input accessory boards to achieve optimal time resolution, and increasing the readout rate to the highest rate achievable by the RFSOC-FPGA. 
To further enhance our readiness for deployment and scability, we plan to work on the design of mechanical packaging that will physically integrate the RFSOC-FPGA with our entangled photon pair source into a rack-mountable form factor.

The successful implementation of these next steps will help to realize our vision for future rack-mounted quantum network nodes to be fully controllable through a single FPGA.

\section*{Acknowledgment}
This work is partially funded by the Department of Energy Advanced Scientific Computing Research Transparent Optical Quantum Networks for Distributed Science program, IEQNET Grant. 
This work is also partially funded by the Department of Energy BES HEADS-QON Grant No. DE-SC0020376 on transduction relevant research for future quantum teleportation systems and communications. 
R.V., L.N., M.S. and S.X. acknowledge partial and S.D. full support from the Alliance for Quantum Technologies’ (AQT) Intelligent Quantum Networks and Technologies (IN-Q-NET) research program that supports the FQNET, CQNET, IEQNET
and other research quantum networking and communications testbeds. 
R.V., L.N., M.S., and S.X. acknowledge partial support from the U.S. Department of Energy, Office of Science, High Energy Physics, QuantISED
program grant, under award number DE-SC0019219. 
L.S., and G.C. acknowledge support from Fermi Research Alliance, LLC under Contract No. DE-AC02- 07CH11359 with the U.S. Department of Energy, Office of Science, Office of High Energy Physics, with support from its QuantISED program and from National Quantum Information Science Research Centers, Quantum Science Center.
Part of the research was carried out at the Jet Propulsion Laboratory, California Institute of Technology, under a contract with the National Aeronautics and Space Administration (80NM0018D0004). 
We thank Jason Trevor (Caltech Lauritsen Laboratory for High Energy Physics and INQNET Laboratory for QST), Vikas Anant (PhotonSpot), Aaron Miller (Quantum Opus), Inder Monga and his ESNET and QUANT-NET groups at LBNL, the groups of Daniel Oblak and Christoph Simon at the University of Calgary, the group of Marko Loncar at Harvard, Artur Apresyan and the HL-LHC USCMS-MTD Fermilab group; Marco Colangelo (MIT); 
We acknowledge the enthusiastic support of the Kavli Foundation on funding QS\&T
workshops and events and the Brinson Foundation support for students working at FQNET and CQNET.




\bibliographystyle{IEEEtran}
\raggedright
\bibliography{Bibliography}


\end{document}